# Exploring The Contribution of Innate Immune Cells to Breast Cancer Immunotherapy

Spatial Analysis of Tumor-Infiltrating Immune Cells with Multiplex RNAscope

Grace Sun[1], Dr. Sandip Patel[2]
[1]The Bishop's School, La Jolla, California
[2]Moores Cancer Center of the University of California, San Diego, California

## Abstract:

Breast cancer is the leading type of cancer in women. About 10-15% of breast cancers are triple-negative breast cancer (TNBC), a subtype with the worst prognosis. Due to the lack of estrogen, progesterone and HER2 receptor expression, chemotherapies have been the standard of care for decades. Immunotherapy has emerged as promising for TNBC treatment. In 2020, the Food and Drug Administration (FDA) granted approval to pembrolizumab in combination with chemotherapy for patients with advanced triple-negative breast cancer. However, only a subgroup of advanced TNBC patients live longer whose tumors have a PD-L1 Combined Positive Score of at least 10 (CPS≥10). There is still an unmet medical need to provide alternative treatment for the rest of patients.

Interestingly, a few of patients at UCSD Moores Cancer Center were found to have had excellent responses to pembrolizumab despite low CPS scores (termed Elite Responders). The hypothesis of this project is that there may be an alternative immune response mechanism and/or crosstalk happening between the innate and adaptive immune systems, especially in Natural Killer Cells and Macrophages, that contributed to this unexpected excellent response. Our procedure used ACDBio RNAscope Multiplex Fluorescence v2 method to spatially analyze innate immune cells (Natural Killer cells and macrophages) and adaptive immune cells (T-cells) in the Tumor Micro Environment. Our data demonstrated increased tumor infiltration of innate immune cells (macrophage and Natural Killer cells) in the Elite Responders. This conclusion indicated the joint effort of two immune systems (innate and adaptive) which eventually led to increased survival.

## Background:

Triple-negative breast cancer (TNBC) accounts for 10%-15% of all breast cancers. It is an aggressive type of breast cancer that does not express Estrogen Receptors (ER), Progesterone Receptors (PR), or normal Human Epidermal growth factor Receptor type 2 (HER2)[1]. TNBC patients have the poorest prognosis due to the lack of targeted therapeutics. TNBC patients have a lower 5-year survival rate (62.1% and 80.8% for TNBC and non-TNBC patients), a higher mortality rate (42.2% in TNBC vs 28% in non-TNBC subtypes) respectively[2,3]. TNBC has high risk of relapse as commonly seen in women younger than age 40, who are of African American descent, or who have a BRCA1 mutation. Visceral metastasis formation is frequent and involves the lung, bone, and brain.

Chemotherapy is the current standard-of-care treatment of this disease. Although TNBCs are highly sensitive to chemotherapy, the frequent occurrence of relapse enforces the search and evaluation of novel therapeutic approaches, such as immunotherapy.

**Breakthrough Treatment Immunotherapy.**
In 2020 the Food and Drug Administration (FDA) approved the combination therapy of pembrolizumab with chemotherapy for patients with advanced Triple-Negative Breast cancer whose tumors had relatively high levels of the PD-L1 protein—a PD-L1 Combined Positive Score of at least 10 (CPS≥10). CPS is essentially a measure of the extent to which cells in a tumor produce PD-L1, the immune checkpoint protein that pembrolizumab targets[4].

Pembrolizumab is a high-affinity, highly selective antibody against PD-1. The programmed death receptor 1 (PD-1) is an inhibitory immune checkpoint receptor expressed on activated T cells, B cells, and natural killer cells. PD-L1, a PD-1 ligand, is an immunosuppressive signal. By blocking immune checkpoints, pembrolizumab and other immune checkpoint inhibitors unleash the immune system against cancer cells. Pembrolizumab is not a silver bullet for advanced TNBC in that only 50% of the patients express PD-L1.

The Human Immune System consists of several major types of lymphocytes including T-cells, B-cells, macrophages, and NK cells. T-cells and B-cells are members of the adaptive immune system that is triggered by the detection of foreign antigens in the blood, macrophage and NK cells belong to the innate immune system[5].

NK cells are the third largest lymphocyte in the blood. NK cells could eliminate tumor cells without antigen-specific cell surface receptors, making them ideal in killing pathogen-infected cells. NK cells use a different mechanism to fight cancers. They are primed to fight any unnaturally altered cells. It is because of this search and destroy mode of action that NK cells are often referred to as the "first line of defense" with regards to the adaptive immune system[6]. NK cells cross-talking among immune cells also play a regulatory control in mediating the anti-tumor adaptive immunity of T- and B-cells[7].

Tumor associated macrophages (TAMs) are long-lived phagocyte cells that not only play a role in phagocytosis but are also involved in antigen presentation to T cells[8]. TAMs can be typically divided into M1 (proinflammatory, tumor killing) and M2 (anti-inflammatory, tumor promoting) types. TNBC release soluble factors such as CSF-1 and extracellular vesicles (Evs) that promote macrophage differentiation into M1 which led to a better prognosis[9].

**Study Rational:** Current TNBC immunotherapies target the PD-L1 and PD-1 axis, which will harness the adaptive immune system to fight cancer. There have been limited studies on the innate immune system in TNBC. A few patients at UCSD Moores Cancer Center displayed excellent responses to pembrolizumab despite low CPS score, and they lived for a considerably longer time than the regular TNBC patients, indicating alternative response mechanisms and/or contributions from other immune cells.

## Materials/Methods and Objectives:

In this study, we analyzed cancer samples from the Elite Responders to explore the roles of macrophage, Natural Killer cells and T cells in Tumor Micro Environments using target-specific probes and RNAscope Multiplex Fluorescent Assay V2 (Figure 1).

## Samples:

Formalin-Fixed Paraffin Embedded (FFPE) Triple Negative Breast cancer biopsies from patients who received Pembrolizumab alone or in combination with chemotherapy, were retrieved from the archives of UCSD Moores Cancer Center biorepository with institutional review board approval (IRB# 15- 0348). Of the 12 samples studied, 6 are from Elite responders who received pembrolizumab longer than 12 months, 4 samples are from patients who didn't respond to Pembrolizumab, and 1 sample is from a patient with high PD-L1 and responded to Pembrolizumab. 8-10 serial sections were collected from each tissue block.

## Assay, fluorophores, and RNA probes

Serial sections of slides were subjected to gene expression analysis by applying RNAscope Multiplex Fluorescence Assay V2 (Part ID: 323100, 323120) to visualize 4 targets simultaneously on FFPE tissues (Figure 1). These included cell specific markers for tumor cells (Pan-CK), Natural Killer cells (NCR1), macrophage (CD68) and T-cells (CD3). Different fluorophores are assigned to the four channels in accordance with the ACD Vivid dyes (Part ID: 323271, 323272 and 323273) and Akoya Biosciences Opal™ polaris 780 dye (FP1501001KT) of that channel and emit a different color upon light excitation (520nm green, 570nm red, 650nm purple and 780nm cyan). ACD RNA probes include Hs-panCK (Part ID: 404751), Hs-NCR1-C2 (Part ID: 312651-C2), Hs-CD68-C3 (Part ID: 560591-C3), Hs-CD3G-C4 (Part ID: 586341-C4), Hs-CD3D-C4(Part ID: 599391-C4), Hs-CD3E-C4 (Part ID: 553971- C4).

## Imaging and Quantification:

The slides were scanned with a Zeiss Axioscan Z.1 epifluorescence slide scanner with a Colibri 7 light source and a 20x (0.8 NA) dry objective. The images were analyzed with QuPath to count the total number of cells in a field of view based on DAPI staining, Tumor cells, NKs, macrophages and T cells were reported as percentage of total number of cells normalized by the number of Pan-CK positive cells.

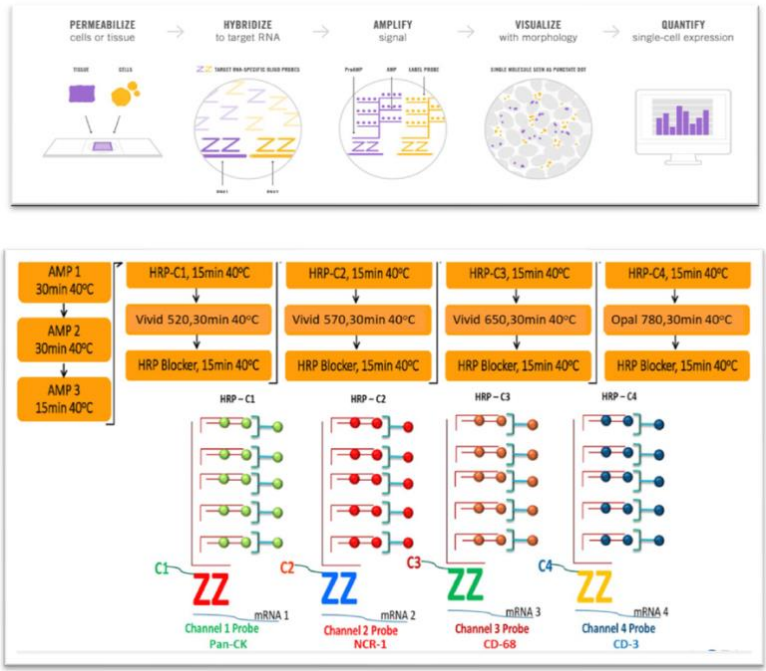

**Figure 1**: Workflow of RNAscope Multiplex Fluorescent V2 assay.

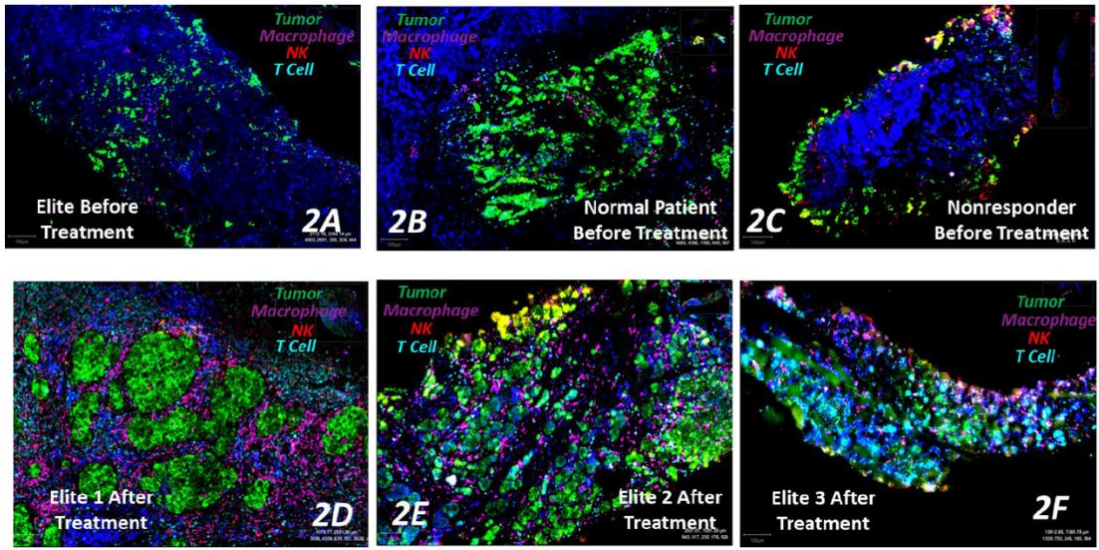

**Figure 2**: Infiltration of macrophage, NK and T cells in Triple Negative Breast Cancer Tumor Micro Environment Visualized by RNAscope Multiplex Fluorescent V2 assay: using a combination of specific marker probes, tumor cells (Pan-CK, green), NK cells (NCR1, red), macrophage (CD68, purple) and T-cells (CD3, cyan) were detected in TNBC samples. Before Pembrolizumab treatment, Elite responder (2A), normal responder (2B) and non-responder (2C) express low levels of macrophage, NKs. Normal responder has higher level of T cells than Elite responder and non-responder (Figure 3). After pembrolizumab treatment, Elite responders demonstrated significant infiltration of macrophages, NKs and T-cells in the TME (2D, 2E, 2F).

| % | NonResponder Before Treatment | Normal Responder Before Treatment | Elite Before Treatment | Elite After Treatment |
|---|---|---|---|---|
| Macrophage | 0.24 | 0.01 | 0.18 | 2.6 |
| NK cells | 0.021 | 0.038 | 0.01 | 0.17 |
| CD3+ T Cell | 0.018 | 1.1 | 0.01 | 5.9 |

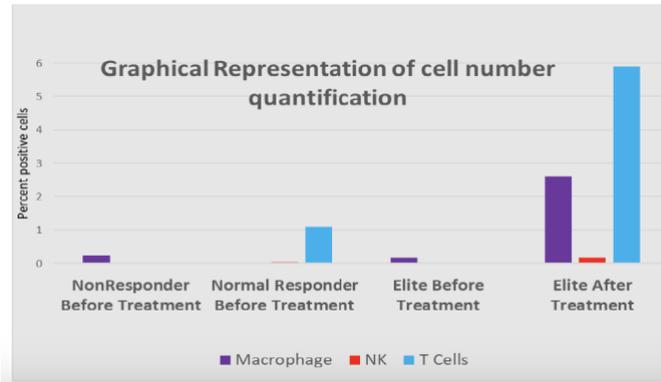

**Figure 3**: Quantification of macrophage, NKs and T cells in Triple Negative Breast Cancer Tumor Micro Environment. Total number of cells in a field of view based on DAPI staining; macrophages, NKs, and T cells were reported as percentage of total number of cells normalized to number of Pan-CK positive cells.

### Results:

RNAscope multiplex Fluorescence assay revealed expression of all four targets (Pan-CK, NKs, macrophage and T cells) with high sensitivity and specificity at single cell level. We were able to successfully visualize spatial relationship between innate and adaptive immune cells in TME when assessing response to the immunotherapy pembrolizumab.

Before the start of pembrolizumab, samples collected from Elite responder (Fig. 2A, n=1), normal responder with high PD-L1 (Fig. 2B, n=1) and non-responder (Fig.2C, n=4) displayed low levels of NKs and macrophage (Fig. 3). However, the normal responder displayed infiltration of T cells in TME (co-localization of Pan-CK and CD3, data not shown). This patient eventually benefited from pembrolizumab, which is expected.

After the start of pembrolizumab, Elite Responders (Fig. 2D, 2E, 2F, n=5) displayed significant increase of all three cell populations. While macrophage tend to overlap with T cells and NKs in TME and adjacent areas indicating a potential crosstalk between innate and adaptive immune cells, NKs were also observed in tumors (co-localization of Pan-CK and NCR-1, data not shown).

The trend is confirmed by cell number quantification in tumor areas selected by the pathologist, analyzed by QuPath (version 0.3.2) and reported as percentage of cell count normalized by Pan-CK positive cell count. Elite responders have an average of 2.6% macrophage, 5.9% T-cells and 0.17% NKs, a significant increase comparing to the Elite responder before therapy start (0.18% macrophage, 0.01% T-cells and 0.01% NKs). Due to the small sample size and lack of access to historical tissue blocks as the true control, we would like to carefully report the observation as a trend.

## Conclusions and Discussions:

Adding immunotherapy drug pembrolizumab to chemotherapy helped some advanced triple negative breast cancer patients live longer but only for those with high levels of PD-L1 proteins (CPS≥10). However, more than half of the TNBC patients have PD-L1 combined positive scores of less than 10. Our study identified a few elite responders to pembrolizumab who have low CPS scores, and spatial analysis of immune cell infiltration of Tumor Micro Environment showed significant involvement of the innate immune system after the start of Pembrolizumab. These findings demonstrate that Elite Responders may use a harmonious mechanism to build antitumor immunity, harnessing both innate and adaptive immune systems to fight TNBC. Our findings echo with recent reports that even if the adaptive immune system is compromised[10] or the function of T cells cannot be fully recovered by PD-1 inhibitors under specific circumstances[11], PD-1/PD-L1 antagonisms can still increase antitumor efficacy.

Tumors are infiltrated by different populations of immune cells including macrophages. Tumor-associated macrophages (TAMs) are heterogenous with M1 (proinflammatory) and M2 (pro-tumor growth) populations. There have been recent reports that M1 dominating TNBC create proinflammatory environment with increased infiltration of T lymphocytes and NK cells and a better prognosis[12,13]. Based on our observation of macrophage involvement in Elite responders and other reports that anti-PD-1 or PD-L1 immune checkpoint blockade induced an M1 macrophage polarization/repolarization[14,15] which lead to enhanced antineoplastic effect, we propose that future studies should focus on the role of M1 subpopulation of macrophage and NKs in TNBC.

In conclusion, my study established a quick, sensitive, and practical methodology of using multiplexed RNAscope to spatially characterize four protein/gene markers and build a digital profile of immune cells (innate and adaptive) in tumor microenvironment, which will help to identify TNBC patients who are more likely to benefit from pembrolizumab. It will be used to molecularly guide the Region of Interest selection for GeoMx Digital Spatial Profiler (DSP) that will perform high plex RNA profiling in cancer patients.

Another implication of my finding is that it gives new hope to the other 50% of the advanced TNBC patients with low CPS scores that addition of macrophage and NK cells to their treatment regimen may lead to longer survival. Both are potential targets for future therapeutic development.

## Acknowledgments:

We want to thank Dr. Oluwole Fadare at the UCSD Division of Anatomic Pathology for his pathology support, Dr. Sara McArdle and Dr. Zbigniew Mikulski at La Jolla Institute for Immunology for the imaging and data analysis support, Chandra Inglis and Lisa Kim for the histology and general lab support.

Compartment Remodeling during Successful Immune-Checkpoint Cancer Therapy. Cell 175, 1014-1030 e1019 (2018)